\documentclass[%
reprint,
 amsmath,amssymb,
aps,
pra,
twocolumn,
showkeys,
]{revtex4-1}
\usepackage{graphicx}% Include figure files
\usepackage{dcolumn}% Align table columns on decimal point
\usepackage{bm}% bold math
\usepackage[hidelinks]{hyperref}% add hypertext capabilities
\hypersetup{colorlinks = true,
  urlcolor = blue,
  linkcolor = blue,
  citecolor = blue
}
\usepackage{caption}
\usepackage{subcaption}
\usepackage{url}
\usepackage[normalem]{ulem}

\begin{document}

\preprint{APS/123-QED}

\title{Fractional integrodifferential equations and (anti-)hermiticity of time in a spacetime-symmetric extension of nonrelativistic Quantum Mechanics  }

\author{Arlans J. S. de Lara}
 \email{arlansslara@gmail.com}
 \author{Marcus W. Beims}%
  \email{mbeims@fisica.ufpr.br}
\affiliation{%
 Departamento de Física, Universidade Federal do Paraná, 81531-990, Curitiba, PR, Brazil
}%

\date{\today}% It is always \today, today,

\begin{abstract}
  Time continues to be an intriguing physical property in the modern era. On the one hand, we have the Classical and Relativistic notion of time, where space and time have the same hierarchy, which is essential in describing events in spacetime. On the other hand, in Quantum Mechanics time appears as a classical parameter, meaning that it does not have an uncertainty relation with its canonical conjugate. In this work, we use a recent proposed spacetime-symmetric formalism~\href{https://doi.org/10.1103/PhysRevA.95.032133}{[Phys.~Rev.~A {\bf 95}, 032133 (2017)]} that tries to solve the unbalance in nonrelativistic Quantum Mechanics by extending the usual Hilbert space. The time parameter $t$ and the position operator $\hat{X}$ in one subspace, and the position parameter $x$ and time operator $\mathbb{T}$ in the other subspace. Time as an operator is better suitable for describing tunnelling processes. We then solve the novel $1/2$-fractional integrodifferential equation for a particle subjected to strong and weak potential limits and obtain an analytical expression for the tunnelling time through a rectangular barrier. We compare to previous works, obtaining pure imaginary times for energies below the barrier and a fast-decaying imaginary part for energies above the barrier, indicating the anti-hermiticity of the time operator for tunnelling times. We also show that the expected time of arrival in the tunnelling problem has the form of an energy average of the classical times of arrival plus a quantum contribution.
\end{abstract}


\maketitle

%\tableofcontents
%==================================================================================================================
\section{\label{Introduction}Introduction}
Time in Quantum Mechanics (QM) has always been a point of discussion \cite{bohm61,razavy67,razavy69,allcock69-1,allcock69-2,allcock69-3,kijo74, aguiar97,briggs1, briggs2, briggs3,hoge10,secao122}, mostly in connection of the time of arrival in QM \cite{tate96,muga97,muga98,muga99,muga00,andersontime,galapon04,galapon05-1,galapon05-2,secao122,galapon06,galapon08,galapon09,galapon18,galapon22}
, including for an unification of QM and General Relativity~\cite{rovelli1991, dittrich2006, dittrich2007, bojowald2011, malkiewicz2017, giacomini2019, montevideointerpretation, bojowald2021, singh, czelusta2022}. Contrary to what happens in Relativity, where spacetime is a single entity~\cite{specialrelativity}, position and time are different kinds of numbers: while the position is a q-number, time is a c-number~\cite{cohen, sakurai, merzbacher}. This hierarchy incompatibility between said quantities has led physicists to search for ways to include a time operator in QM. Though Pauli argued~\cite{nogopauli} that a bounded-from-below Hamiltonian is incompatible with a time operator canonically conjugate to it (both the Hamiltonian and the time operator must possess completely continuous spectra spanning the entire real line), there are ways to overcome it (see, for example \cite{bohm61,muga98,galapon04,galapon05-1}). One of the most famous related works is the Page and Wootters (PaW) mechanism, together with its recent interpretations~\cite{paw, giovannetti, macconesacha, mendespinto}, in which the universe is in a stationary state, consistent with a Wheeler-DeWitt equation~\cite{andersontime, dewitt}, and the apparent dynamical evolution that systems undergo is relative to the degrees of freedom of the rest of the universe, that acts like a clock.

Not considering time as an operator, interpretations of the relation between time and energy were also made by Mandelstam and Tamm~\cite{mandelstamtamm} and Margolus and Levitin~\cite{margoluslevitin}. Any $\Delta t$ appearing in those works must be interpreted as a time interval, not an operator uncertainty. In both cases, $\Delta t$ is considered the smallest time interval for a system to evolve into an orthogonal state. However, in the former, the system has an energy \emph{spread} $\Delta E$, which bonds the interval by $\Delta t \Delta E \sim \hbar$. In contrast, in the latter, the system has an \emph{average} energy $\left< E \right>$, bounding the interval from below by $\pi \hbar/ 2 \left< E \right>$.

Using the idea of quantum events~\cite{dias, fadelmaccone}, it is possible to give meaning to the usual time-energy ``uncertainty'' relations and relate the uncertainty of a quantum measurement of time to its energy uncertainty. By requiring consistency with the way that time enters the fundamental laws of Physics, one can also draw a picture where it is shown that there is only one time: both classical and quantum times are manifestations of entanglement~\cite{foticoppo}.

Höhn et al. show that there is an equivalence between the relational quantum dynamics, (a) the relational observables in the clock-neutral picture of Dirac quantization, (b) the PaW mechanism and (c) the relational Heisenberg picture obtained via symmetry reduction using quantum reduction maps~\cite{hoen1, hoen2}.

The spacetime-symmetric formalism~\cite{diasparisio, diasparisioexp} that we present and use in this paper uses similar ideas to the PaW mechanism. The system has a new Hilbert space with a \emph{operator} time, implying an extended state for the system that depends on variables both in the usual position Hilbert space and variables in the new temporal Hilbert space. One key difference to the PaW formalism is that this new Hilbert space is as intrinsic to the system as the position Hilbert space, and no auxiliary systems are required. This provides a clear interpretation of the time-energy uncertainty relation and different types of experiments, where predictions of the positions of particles, the time of arrival, or both, can be obtained. It is a natural subject then to examine the tunnelling times in the spacetime-symmetric formalism. This formalism has shown promising results compared to the Büttiker-Landauer and Phase-Time~\cite{BL1, secao122, BL3, BL4} approaches to time-of-arrival problems and is the main reason we use this formulation. Worth calling attention to is that $1/2$-fractional time derivatives and integrations appear in the equations to be solved.

Our goal in this paper is to study weak and strong potential barriers, providing connection formulas for the wave functions in the \textit{extended Hilbert space}, and examine results for the tunnelling times by comparing them with tunnelling times obtained using the usual QM formalism. The tunnelling time through a barrier is an old problem which goes back to MacColl \cite{coll32}. The exact definition, applicability, and measure of tunnelling times change according to the circumstances and interest: dwell times, arrival times, asymptotic phase times, delay times, and jump times, among others. It is impossible to furnish a fair review of all these works here, but we refer the readers to \cite{hauge89,landauer94,secao122}. Some particular cases will be mentioned later for comparison.

This paper is organized as follows. While in Sec.~\ref{stsdiasparisio}, we summarize the spacetime-symmetric formalism, in Sec.~\ref{results}, we obtain and solve the approximated $1/2-$fractional equations for the weak and strong potential limits. In Sec.~\ref{aplicacao}, we apply the results obtained in Sec.~\ref{results} to the tunnelling time and show that the results obtained are pure imaginary for energies below the barrier, and with a fast decaying imaginary part for energies above. We obtain an average of classical times-of-flight plus a quantum correction up to the first order. Final comments are presented in Sec.~\ref{conclusion}.

%==================================================================================================================
\section{\label{stsdiasparisio}The spacetime-symmetric formalism}
We begin our discussion by revisiting the formalism used in this paper. The spacetime-symmetric formalism (hereafter the STS formalism) of QM proposed by Dias and Parisio~\cite{diasparisio, diasparisioexp} uses a similar idea to the PaW mechanism~\cite{paw,giovannetti}, in which the entire Hilbert space is divided into one subspace that refers to the \emph{system of interest}, and another one, that refers to the \emph{clock}. The main difference  between the two is that the complete Hilbert space in the STS formalism,
\begin{eqnarray}
\mathcal{H} = \mathcal{H}_{\text{\tiny pos}} \otimes \mathcal{H}_{\text{\tiny time}},
\end{eqnarray}
(here $\mathcal{H}_{\text{\tiny pos}}$ is the usual Hilbert space of QM and $\mathcal{H}_{\text{\tiny time}}$ is a temporal \emph{extension} of the regular theory) refers entirely to the system: $\mathcal{H}_{\text{\tiny time}}$ is as intrinsic to the system as $\mathcal{H}_{\text{\tiny pos}}$ in this approach.

In this new space we define the time operator $\mathbb{T}$ with eigenkets $\left|t \right>$ as
\begin{eqnarray}
\mathbb{T} \left| t \right> = t \left| t \right>,
\end{eqnarray}
where $t$ is the eigenvalue associated with $\left|t \right>$. The set of eigenkets 
$\{\left| t \right>\}$ resolve an identity $\mathbb{I} = \int_{-\infty}^{\infty} \mathrm{d} t \, \left| t \middle> \middle< t \right|$. We then define the energy operator $\mathbb{H}$ through the commutation relation~\footnote{The motivation for the minus signal in this definition is similar to the position-momentum relation in the usual Quantum Mechanics: the application of the momentum operator on a plane-wave $\exp\left(-i(Et - px)/\hbar \right)$ gives $+p$. In contrast, a similar energy operator should give the result $E$. Since $p$ is written as $-i \partial_x /\hbar$ in the position representation, the negative signal of the commutator allows us to write the energy operator as $+ i \partial_t /\hbar$ in the new Hilbert space.}
\begin{eqnarray}
\left[\mathbb{T},\mathbb{H}\right] = - i \hbar,
\end{eqnarray}
which gives us naturally the time-energy uncertainty relation $\Delta \mathbb{T} \Delta \mathbb{H} \geq \hbar/2$. We want to emphasize that, since the STS formalism considers an extension of the Hilbert space of the system of interest, this uncertainty relation relates the energy \emph{of the system} and a time operator that acts \emph{on the system}. This does not happen when you consider, for example, the PaW mechanism, where an auxiliary system takes the role of a clock~\cite{paw}. The price paid for this in the STS formalism is that we do not have the commutation relation $\left[x, \mathbb{P}\right] \propto i \hbar$ since in the new Hilbert space $x$ is a classical parameter.

The complete state of the system is given by
\begin{eqnarray}
\left| \middle| \Psi \right> = \int \int \mathrm{d} x \, \mathrm{d} t \, \Psi \left(x \,\& \,t\right) \left| x\right> \otimes \left| t \right>.
\end{eqnarray}
The double ket notation indicates that this state belongs to both Hilbert spaces. The way we write the argument of $\Psi \left(x \,\&\, t \right) \equiv \left(\left<x \right| \otimes \left< t \right| \right) \left| \middle| \Psi \right>$ is such as to reminds us that, in this formalism, position and time will be on equal footing, but with some caveats that will be made clear later.

The square modulus of $\Psi \left(x \,\& \,t \right)$ is related to the wave functions in their respective spaces as
\begin{eqnarray}
  \mathcal{P}(x,t)\, \mathrm{d}x\, \mathrm{d} t  &=& \left| \Psi \left( x \, \& \, t \right) \right|^2 \, \mathrm{d}x\, \mathrm{d} t \nonumber\\
  &=& \left|\psi(x|t) \right|^2 f(t)\, \mathrm{d}x\, \mathrm{d}t \nonumber\\
&=& \left|\phi(t|x) \right|^2 g(x)\, \mathrm{d}x\, \mathrm{d}t,
\end{eqnarray}
where $\mathcal{P}(x,t)\, \mathrm{d}x\, \mathrm{d} t$ is the probability of finding a particle in the length interval $[x, x+\mathrm{d}x]$ and the time interval $[t, t+\mathrm{d} t]$. The notation in $\psi(x|t)$, the usual wave function, means that $\left|\psi(x|t) \right|^2 \mathrm{d}x$ is the probability of finding the particle in the length interval $[x, x+\mathrm{d}x]$ \emph{given that} the clock reads $t$.  Through Bayes' rule~\cite{bayesrule}, it implies that $f(t)\, \mathrm{d}t$ is the probability of finding the particle in the time interval $[t,t+\mathrm{d} t]$; analogously for $\phi(t|x)$ and $g(x)$. The functions $f(t)$ and $g(x)$ cannot be given through the equations of the systems alone; these depend on the type of experiment, the settings of the laboratory, etc~\cite{diasparisio}.

The STS formalism, then, tells us that if the experiment does not require predictions on time (for instance, the fringes at the end of the run of a double-slit experiment, all we need is the usual wave function $\psi(x|t)$. If we need only predictions of time (e.g. tunnelling through a potential barrier), all that is required is $\phi(t|x)$. In cases where predictions of both position \emph{and} time are needed, the complete wave function $\Psi(x \,\& \,t)$ should be used.

The ``dynamics" in $\mathcal{H}_{\text{\tiny time}}$ is given by
\begin{eqnarray}
\mathbb{P} \left| \phi(x) \right> = - i \hbar \frac{\partial}{\partial x} \left| \phi(x) \right>,
\end{eqnarray}
with $\mathbb{P}$, the Momentum operator in $\mathcal{H}_{\text{\tiny time}}$, defined as
\begin{eqnarray}
\mathbb{P} \equiv \sigma_z \sqrt{2m \left(\mathbb{H} - V(x,\mathbb{T}) \right)},
\end{eqnarray}
$\sigma_z = \text{diag}(1,-1)$. When projected on $\left< t \right|$, this leads us to
\begin{eqnarray}
  \label{eqprojetada}
\sigma_z \sqrt{2m \left(i \hbar \frac{\partial}{\partial t} - V(x,t) \right)} \phi(t|x) = - i \hbar  \frac{\partial}{\partial x} \phi(t|x),
\end{eqnarray}
where $\phi(t|x) = \left< t \middle| \phi(x) \right>$. The quotation marks in ``dynamics" mean that $\mathbb{P}$ generates variations in the (classical) parameter position. Compare this to the usual QM, where the Hamiltonian is the generator of variations in the (classical) parameter time. Because of the presence of $\sigma_z$, $\phi(t|x)$ is a pseudo-spinor with components
\begin{eqnarray}
\phi(x|t) = \left(
\begin{array}{c}
\phi^+(x|t)\\
\phi^-(x|t)
\end{array}\right).
\end{eqnarray}
As such, the square modulus then is given by $\left|\phi(t|x)\right|^2 = \phi^{\dagger}(t|x) \phi(t|x)$.

A comment about this formalism is necessary. In the usual QM, $\hat{X}$, $\hat{P}$ and $\hat{H}$ are operators acting on $\mathcal{H}_{\text{\tiny pos}}$, and $t$ is a classical parameter. The point of the STS formalism is that the extended Hilbert space \emph{symmetrizes} operators and parameters: on the one hand, we have position and momentum as operators, and the generator of the dynamics, the Hamiltonian, as a function of these two, with the label $t$ acting as a parameter; on the other hand, we have \emph{time} and \emph{energy} as operators, and the \emph{momentum} is the generator of the ``dynamics'', while still having a classical parameter: in this case, the position $x$ of the particle. This is why, for time-of-flight experiments, all we need is $\phi(t|x)$: the measuring devices are classical objects, meaning that we have, in principle, arbitrary precision of \emph{where} the device is located. Then $x$ has to act as a classical parameter.

Of course, if we consider the detectors to be lightweight and behave quantum mechanically~\cite{renatosavi2017}, the uncertainty in the detector's position would be significant, and we would not be able to apply only the STS formalism. Since this will not be the case in the present work, we do not have to worry.

\subsection{Expectation values in the spacetime-symmetric formalism}
In the usual QM formalism, experimental results from measuring a quantity that has an operator $\hat{A}$ related to it are compared to the expectation value via
\begin{eqnarray}
\langle\hat{A}\rangle(t) = \frac{\left< \psi(t) \right| \hat{A} \left| \psi(t) \right>}{\left< \psi(t) \middle| \psi(t) \right>},
\end{eqnarray}
which corresponds to averaging measurements of $\hat{A}$ in an ensemble of identically prepared systems, given that we measure at time $t$. We usually do not write the denominator because the wave function is normalized to the unit and its normalization is a constant:
\begin{eqnarray}
i \hbar \frac{\partial}{\partial t} \left< \psi(t) \middle| \psi(t) \right> = \left< \psi(t) \right| \left(\hat{H} - \hat{H}^{\dagger}\right) \left| \psi(t) \right> = 0,
\end{eqnarray}
because of the hermiticity of the Hamiltonian~\cite{sakurai, cohen, merzbacher}.

Now, consider the expectation value in $\mathcal{H}_{\text{\tiny time}}$. We have, as in $\mathcal{H}_{\text{\tiny pos}}$,
\begin{eqnarray}
  \label{valoresperado}
\left< \mathbb{B} \right>(t) = \frac{\left< \phi(x) \right| \mathbb{B} \left| \phi(x) \right>}{\left< \phi(x) \middle| \phi(x) \right>},
\end{eqnarray}
having a similar interpretation of the average of measurements of $\mathbb{B}$, given that the measurement happened at the position $x$. However, in contrast to what happens in $\mathcal{H}_{\text{\tiny pos}}$, the denominator is generally not constant.

The physical interpretation, given by Ref.~\cite{diasparisioexp}, is that in the usual QM, the particle is expected to exist in some position, regardless of the instant of the measurement. This is different in the extended space, in general. Consider the double-slit experiment: there are points in space where the particle never arrives, independent of how long we wait. If we mirror the interpretation, the difference is clear: the particle \emph{should} exists in \emph{some} instant of time, \emph{independent} of the position of the measurement. This does not happen in general; dark regions on the fringes illustrate this. Some regions are forbidden, no matter how long we wait for the particle to arrive. This means that whenever we use the STS expectation values, we have to carry the factor $\left<\phi(x) \middle| \phi(x) \right>$ throughout the calculations.

%==================================================================================================================
\section{\label{results}Weak and strong potential approximations}
To obtain the wave function in the extended space, we need to solve Eq.~(\ref{eqprojetada}). This is difficult because of the appearance of a derivative operator inside the square root. We can, however, consider the two extreme cases of weak and strong potentials, which enables us to obtain approximate equations in these limits that can be applied, for instance, to scattering and tunnelling problems.

%==================================================================================================================
\subsection{\label{weakpotential}Weak potential}
Since the generator of the ``dynamics'' in $\mathcal{H}_{\text{\tiny time}}$ is a function of the operators $\mathbb{T}$ and $\mathbb{H}$, we expand the momentum operator $\mathbb{P}$ in a Taylor series up to first order. For this, we examine the actuation of $\mathbb{H}$ to be greater than that of $V(x,\mathbb{T})$ in some sense. We consider the expected value of these operators to fulfil
\begin{equation}
\left<\phi(x)\right| \mathbb{H} \left| \phi(x) \right> \gg \left<\phi(x)\right| V(x,\mathbb{T}) \left| \phi(x) \right>.
\end{equation}
Thus, the expectation value of the potential energy is small compared to the expectation value of the total energy, meaning that the particle rarely will have significant potential energy. Mathematically,
\begin{align}
    \mathbb{P} &= \sigma_z \sqrt{2 m \left( \mathbb{H} - V(x,\mathbb{T}) \right)}, \nonumber\\
    &\simeq \sigma_z\sqrt{2 m \mathbb{H}} \left[1 - \frac{1}{4} \left(\frac{1}{\mathbb{H}} V(x, \mathbb{T}) + V(x, \mathbb{T})\frac{1}{\mathbb{H}} \right) \right],
  \end{align}
where we used $\sqrt{1 + \lambda} \simeq 1 + \lambda/2$ for sufficiently small $\lambda$, and since $\mathbb{H}$ does not commute with $\mathbb{T}$, we symmetrize the expansion. For simplicity, from now on, we will consider the potential to be independent of time, which gives us
\begin{eqnarray}
  \mathbb{P} \simeq \sigma_z \sqrt{2m} \left[ \mathbb{H}^{1/2} - \frac{1}{2}\frac{V(x)}{\mathbb{H}^{1/2}}  \right].
  \end{eqnarray}
When projected on $\left<t\right|$, the operators $\mathbb{H}^{1/2}$ and $1/\mathbb{H}^{1/2} \equiv \mathbb{H}^{-1/2}$ produces $1/2$-\emph{fractional} derivatives and integrals, which can be defined as the Caputo fractional derivative~\footnote{{In our opinion, the Caputo fractional derivative has an adequate physical interpretation, since its derivative of a constant vanishes. Besides that, the Laplace transform of the Caputo derivative depends only on initial conditions of \emph{integer} order.}} and the Riemmann-Liouville fractional integral, respectively~\cite{oldhamspanier, laskin, mathaisaxena, herrmann, samkokilbas}. Then, we will have
\begin{eqnarray}
    - i \hbar \partial_x \phi(t|x) &=& \sigma_z \sqrt{2m i \hbar} \partial_t^{1/2} \phi(t|x) \nonumber\\
      & &- \sigma_z\sqrt{\frac{ m}{2 i \hbar}} V(x) \partial_t^{-1/2} \phi(t|x).
  \end{eqnarray}
This fractional partial differential equation can be, in principle, solved through different methods, for instance, the Laplace transform of fractional derivatives and integrals~\cite{oldhamspanier, laskin, mathaisaxena, herrmann, samkokilbas}. For now, we will focus on the case of a constant potential $V(x) = V_0$. It is then possible to separate the equation onto temporal and spatial parts if we consider $\phi(t|x) = F(t) G(x)$:
\begin{subequations}
  \begin{align}
\begin{split}
  p\, G(x) &=- \sigma_z i \hbar \partial_x G(x);
\end{split}\\
\begin{split}
   p\, F(t) &= \sqrt{2 m } \left[ \sqrt{i \hbar} \partial_t^{1/2} - \frac{V_0}{2 \sqrt{i \hbar}} \partial_t^{-1/2}\right]F(t),
\end{split}
\end{align}
\end{subequations}
$p$ being the constant of separation, and we made use of the linearity of the fractional derivatives and integrals~\cite{oldhamspanier, laskin, mathaisaxena, herrmann, samkokilbas}. We use the ansatz
\begin{eqnarray}
  \label{ansatzpotencialfraco}
    G^{\pm}(x) &=& \exp \left[ \pm \frac{i}{\hbar} p x \right] \nonumber \\
    F(t) &=& \exp \left[ - i \omega t\right],
\end{eqnarray}
where $G^{\pm}(x)$ are the $\pm$ spatial components of the spinor, together with the fractional derivative property~\cite{diasparisio, oldhamspanier, laskin}
\begin{equation}
  \partial_t^{\alpha} \exp\left[ \beta t \right] = \beta^{\alpha} \exp \left[ \beta t \right],
  \end{equation}
to obtain
 \begin{equation}
  %    \label{energiapotencialfraco}
    p =  \sqrt{2m \hbar \omega}\left(1-\frac{V_0}{2 \hbar \omega} \right),
 \end{equation}
 that is, the first order approximation of a particle with momentum $p = \sqrt{2m (E - V_0)}$ and energy $E = \hbar \omega$. Thus, the momentum in the STS formalism is consistent with known results from Classical Mechanics (CM) and QM, at least in the weak and constant potential approximation.

 Using this approximation, we can solve for $E$ and arrive at
 \begin{equation}
         \label{energiapotencialfraco}
   E = \frac{p^2}{2m} + V_0.
 \end{equation}
 If we apply this to the case $V_0 = 0$, we obtain the solution for the free particle obtained in Ref.~\cite{diasparisio},
\begin{equation}
  \phi^{\pm}(t|x) = \exp\left[- \frac{i}{\hbar} \frac{p^2}{2m} t \pm \frac{i}{\hbar} p x \right],
  \end{equation}
as expected.

%==================================================================================================================
\subsection{Strong potential}
\label{strongpotential}
Considering a Taylor series expansion of the momentum operator with a strong, time-independent potential, we can write
\begin{equation}
  \mathbb{P} \simeq \sigma_z \sqrt{-2 m V(x)} \left[1 - \frac{\mathbb{H}}{2V(x)} \right],
  \end{equation}
which leads us to
\begin{equation}
  \label{equacaofortesemseparar}
  \sqrt{-2 m V(x)} \left[1 - \frac{i \hbar \partial_t}{2V(x)} \right] \phi(t|x) = -\sigma_z i \hbar \partial_x \phi(t|x).
  \end{equation}
Curiously, in the strong potential approximation, the order of the derivatives is the same. Separating this equation enables us to write
\begin{subequations}
  \label{separadaforte}
  \begin{align}
    \begin{split}
      \label{separadafortea}
      i \hbar \partial_t F(t) &= E F(t);
    \end{split}\\
    \begin{split}
      \label{separadaforteb}
      \sigma_z i \hbar \partial_x G(x) & = \sqrt{\frac{-m}{2 V(x)}} \left[E - 2 V(x) \right] G(x),
      \end{split}
    \end{align}
\end{subequations}
where, as before, $\phi(t|x) = F(t) G(x)$, and $E$ is the separation constant. Eq.~(\ref{separadafortea}) is trivial, giving us
\begin{equation}
  \label{fforte}
  F(t) = \exp \left(- \frac{i}{\hbar} E t \right),
\end{equation}
compatible with known results from the usual QM~\cite{sakurai, cohen, merzbacher}. Since we are considering a strong potential, we can rewrite the right-hand side of Eq.~(\ref{separadaforteb}) as
\begin{equation}
  \sqrt{\frac{-m}{2 V(x)}} \left[E - 2 V(x) \right] \simeq - \sqrt{2 m \left(E - V(x)\right)},
  \end{equation}
giving us
\begin{equation}
  G^{\pm}(x) = \exp \left[ \pm \frac{i}{\hbar} \int_{x_0}^{x} \mathrm{d} x' \, \sqrt{2 m \left(E - V(x') \right)}\right],
  \end{equation}
where $x_0$ depends on the boundary conditions. The constant potential is trivial and gives us, up to a multiplication constant,
\begin{equation}
  \label{ansatzpotencialforte}
 G(x)  = \exp\left[ \pm \frac{i}{\hbar} p x \right],
\end{equation}
with $p = \sqrt{2m (E - V_0)} \in \mathbb{C}$ being the momentum of the system, which again coincides with the CM and QM momenta relations, subject to a constant potential with intensity  $V_0$.

Notice that we obtained the relation $p = \sqrt{2m (E - V_0)}$ without any ad hoc hypothesis; the momentum was obtained through the dynamics of the STS formalism, as opposed to Ref.~\cite{diasparisioexp}. Our results confirm their findings.

%==================================================================================================================
\section{\label{aplicacao}Results}

%==================================================================================================================
\subsection{Toy model: rectangular potential barrier}
The toy model we use for our main result is the textbook potential barrier:
\begin{equation}
  V(x) = \left\{
  \begin{array}{lcl}
    V_0 = \text{const}, &\quad& 0 < x < L;\\
    0,& &\text{everywhere else}.
    \end{array} \right.
  \end{equation}
$V_0$ is such that we can use the strong potential limit of Sec.~\ref{strongpotential} for this region, and $L$ is the length of the barrier.

We want the wave function to be continuous in the interfaces $x = 0$ and $x = L$ for all instants of time, following the same principles as in the usual QM~\cite{sakurai, cohen, merzbacher}. Since $F(t)$ has the same form for all regions, the temporal connection is trivial and implies that the energies $E = \hbar \omega$ must be equal in all regions. Then, for the spatial part, we consider
\begin{eqnarray}
  \label{parteespacialsemconexao}
  G^{\pm}(x) = \left\{
  \begin{array}{lll}
\displaystyle    A_1^{\pm} \exp \left[\pm \frac{i}{\hbar} p_1 x \right], & \quad & x < 0;\\[10pt]
\displaystyle    A_2^{\pm} \exp \left[\pm \frac{i}{\hbar} p_2 x \right], & \quad & 0< x < L;\\[10pt]
\displaystyle        A_3^{\pm} \exp \left[\pm \frac{i}{\hbar} p_1 x \right], & \quad & L < x,
    \end{array} \right.
  \end{eqnarray}
with
\begin{eqnarray}
    p_1 &=& \sqrt{2m E},\nonumber\\
    p_2 &=& \sqrt{2m (E - V_0)}.
  \end{eqnarray}
Connecting the wave function at the interfaces, we have
\begin{eqnarray}
  \label{conexaoxl}
    A_2^{\pm} &=& A_1^{\pm},\nonumber\\
    A_3^{\pm} &=& A_1^{\pm} \exp\left[\pm\frac{i}{\hbar} \left(p_2 - p_1 \right)L \right].
    \end{eqnarray}
Combining Eqs.~(\ref{parteespacialsemconexao}) and~(\ref{conexaoxl}), together with $F(t) = \exp \left( - i E t/\hbar \right)$, we have the total wave function for the rectangular barrier.

%==================================================================================================================
\subsection{Tunneling time}
As in~\cite{diasparisioexp}, we define the time of travel (or, in the specific case we want to tackle in this section, tunnelling time) as the difference between the expectation values:
\begin{equation}
  T_{ \text{\tiny STS}} \left(x_{\text{i}} \to x_{\text{f}} \right)  = \left< \mathbb{T} \right> (x_{\text{f}}) - \left< \mathbb{T} \right>(x_{\text{i}}),
\end{equation}
with $ \left< \mathbb{T} \right> (x)$ is given by Eq.~(\ref{valoresperado}).

The solutions that led to Eqs.~(\ref{parteespacialsemconexao}) and (\ref{conexaoxl}) are eigenfunctions of $\mathbb{P}$, with eigenvalues $p = \sqrt{2 m E}$ outside the barrier or $p = \sqrt{2m (E - V_0)}$ inside the barrier.  When we prepare systems for experiments in the usual QM, we generally consider a wave packet, which is a linear combination of eigenfunctions of the Hamiltonian $\hat{H}$. In the same manner, since $\mathbb{P}$ is a linear operator, linear combinations of solutions of Eq.~(\ref{eqprojetada}) are also solutions of the same equation. In this manner, the wave packet is written as
\begin{equation}
  \phi^{\pm}(t|x) = \int \mathrm{d} E \, C^{\pm}_E \exp \left(- \frac{i}{\hbar} E t \right) G^{\pm}(E, x),
  \end{equation}
where the limits must respect the condition of strong and/or weak potential, depending on the region, and $C_E^{\pm}$ is the energy distribution for the wave packet. The discrete case is straightforward. The correct way of writing the wavepacket should be in terms of eigenfunctions and eigenvalues of $\mathbb{P}$. Since we know the relation between $p$ and $E$ (e.g., $p = p(E)$), this is, at heart, just a change of variables in the integration, with the distributions $C_E^{\pm}$ having to change accordingly~\cite{diasparisio}. We also changed the notation from $G^{\pm}(x)$ to $G^{\pm}(E, x)$ to emphasize the energy dependence of the spatial part. We are making an abuse of notation using the same $\phi^{\pm}(t|x)$ as before, but since from now on, we will only work with the wave packet, there should be no confusion.

Using the completeness relation $\int_{-\infty}^{\infty} \mathrm{d} t \, \left| t \middle> \middle< t \right| = \mathbb{I}$, we can write the expectation value of $\mathbb{T}$ as
\begin{eqnarray}
  \label{valoresperadot}
  \left< \mathbb{T} \right> (x) &=& \displaystyle\frac{\left<\phi(x) \right| \mathbb{T} \left| \phi(x) \right>}{\left< \phi(x) \middle| \phi(x) \right>} \nonumber\\
  & =& \displaystyle \frac{\int_{-\infty}^{\infty} \mathrm{d} t \, t\, \rho(t|x)}{\int_{-\infty}^{\infty} \mathrm{d} t \, \rho(t|x)},
  \end{eqnarray}
where 
\begin{eqnarray}
  \rho(t|x) &=& \phi^{\dagger}(t|x)\phi(t|x) \nonumber\\
  &=& \left| \int \mathrm{d} E \,\, C^{+}_E \exp \left(- \frac{i}{\hbar} E t \right) G^{+}(E, x) \right|^2 \nonumber\\
  && + \left| \int \mathrm{d} E \,\, C^{-}_E \exp \left(- \frac{i}{\hbar} E t \right) G^{-}(E, x) \right|^2.
\end{eqnarray}

To calculate the expectation value in Eq.~(\ref{valoresperadot}), we can write the numerator as
\begin{eqnarray}
    N &\equiv& \int_{-\infty}^{\infty} \mathrm{d} t \, t\, \rho(t|x) \nonumber\\
    &=& \sum_{r= \pm} \int_{-\infty}^{\infty} \mathrm{d} t \, t\, \left[\int \mathrm{d} E \, C_E^r F(t) G^r(E,x)\right] \nonumber\\
    && \times \left[\int \mathrm{d} E' \, C_{E'}^r F'(t) G^r(E',x)\right]^*,
  \end{eqnarray}
where the prime denotes that we need to substitute $E \to E'$ in the argument of the second integral. The temporal integral can be rewritten as
\begin{equation}
  \int_{-\infty}^{\infty} \mathrm{d} t \, t \, \exp \left[- \frac{i}{\hbar} \left(E - E' \right)t \right] = - 2 \pi i \hbar^2 \partial_{E'} \delta(E' - E),
  \end{equation}
where we made use of
\begin{equation*}
  t \, \exp \left[- \frac{i}{\hbar} \left(E - E' \right)t \right] = - i \hbar \partial_{E'} \exp \left[- \frac{i}{\hbar} \left(E - E' \right)t \right],
\end{equation*}
and the integral representation of the Dirac delta~\cite{arfkenmethods}
\begin{equation*}
  2 \pi \delta(x - a) = \int_{-\infty}^{\infty} \mathrm{d} p \, \exp \left[ i p (x - a) \right].
\end{equation*}
Then, the numerator becomes
\begin{equation}
  N = 2 \pi i \hbar^2 \sum_{r=\pm} \int \mathrm{d} E \, C_E^r G^r(E,x) \partial_E \left[ C_E^r G^r(E,x) \right]^*.
  \end{equation}
Similarly, we can write the denominator as
\begin{eqnarray}
    D &\equiv& \int_{-\infty}^{\infty} \mathrm{d} t \,\rho(t|x) \nonumber\\
    &=& 2 \pi \hbar \sum_{r =\pm} \int \mathrm{d} E \, \left|C_E^r G^r(E,x) \right|^2,
  \end{eqnarray}
which finally brings us to
\begin{equation}
  \label{valoresperadofinal}
  \left< \mathbb{T} \right> (x) = i \hbar \frac{\sum_{r=\pm} \int \mathrm{d} E \, C_E^r G^r(E,x) \partial_E \left[ C_E^r G^r(E,x) \right]^*}{\sum_{r =\pm} \int \mathrm{d} E \, \left|C_E^r G^r(E,x) \right|^2}.
  \end{equation}
Eq~(\ref{valoresperadofinal}), together with Eqs.~(\ref{parteespacialsemconexao}) and~(\ref{conexaoxl}) allow us to predict tunnelling times and dwell times whenever the potential is sufficiently strong and constant. Because, in principle, we can position the probes with arbitrary precision in this treatment, the time it takes for the particle to tunnel, on average, is given by
\begin{equation}
  \label{travelLto0}
T_{\text{\tiny STS}}(0 \to L) = \left< \mathbb{T} \right> (L) - \left< \mathbb{T} \right> (0)
\end{equation}
for a potential barrier located between $x=0$ and $x=L$~\cite{diasparisioexp}.

\subsection{\label{applicationstrong} Application of tunnelling time: a constant distribution for a wave packet moving to the right inside the barrier}
\begin{figure}[t]
     \centering
     \begin{subfigure}[b]{0.4\textwidth}
         \centering
         \includegraphics[width=\textwidth]{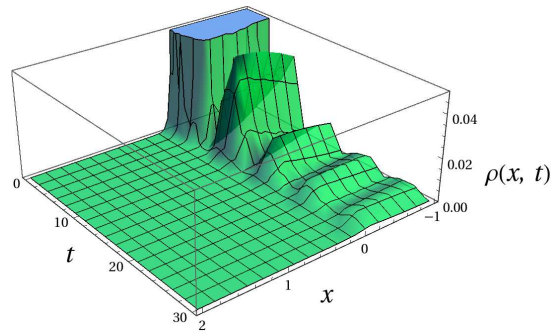}
        \label{phimais}
     \end{subfigure}
%     \hfill
%     \begin{subfigure}[b]{0.4\textwidth}
 %        \centering
  %       \includegraphics[width=\textwidth]{phimenoszoom.eps}
   %      \label{phimenos}
    % \end{subfigure}
%     \hfill
        \caption{$\rho(x|t)$ for $m = \hbar = L = 1$ and $V_0 = 100$, in arbitrary units. $C_E^+ = 1$ and $C_E^- = 0$, meaning that plane wave moves initially from the left ($x=-1$) to the right ($x=2$).}
        \label{figrho}
\end{figure}

For an application of Eq.~(\ref{valoresperadofinal}), we consider the following: $C_E^+ = C = \text{constant}$ and $C_E^- = 0$. This second equality means that, since the expressions in Eq.~(\ref{parteespacialsemconexao}) are plane waves, the wave travelling from right to left on the real $x$-axis is discarded. 
Fig.~\ref{figrho} displays $\rho(t|x)$ for a constant distribution and the rectangular potential barrier between $x=0\to 1$. The plane wave packet moves from left to right (that is, from most negative $x$ to most positive $x$). Notice that since we have a non-normalized wave function, the absolute values can be very high.  As $t$ increases to $5$, reasonable finite values of $\rho(t|x)$ are observed inside the barrier. Then,  a sudden decrease of $\rho(t|x)$ to zero is visible, even for $x<0$, away from the barrier. For increasing times, $\rho(t|x)$ oscillates periodically with diminishing amplitude, which certainly decreases the tunnelling probabilities for larger times.

For the constant distribution case, the expectation value of $\mathbb{T}$ can be simplified as
\begin{equation}
  \left< \mathbb{T} \right> (x) = i \hbar \frac{\int \mathrm{d} E \,  G(E,x) \partial_E \left[G(E,x) \right]^*}{ \int \mathrm{d} E \, \left|G(E,x) \right|^2},
  \end{equation}
where we dropped the $r = \pm$ indexes since we only have the $r = +$ component. The derivative inside the integral requires some attention. The integral in the numerator can be rewritten as
  \begin{eqnarray}
\label{integralnumeradorcomplexo}
    &&\int \mathrm{d} E \,  G(E,x) \partial_E \left[G(E,x) \right]^* \nonumber\\
    &&= \int \mathrm{d} E \,\frac{mi}{\hbar \sqrt{2 m (E - V_0)}} \exp \left[- \frac{2 x}{\hbar} \mbox{Im} \left( \sqrt{2 m (E - V_0)} \right) \right].\nonumber\\
  \end{eqnarray}
This integral is also valid for the expectation value outside the barrier, as will be used in Eq.~(\ref{tempoweakpotential}). Since $E < V_0$ in the region of interest ($0 \leq x \leq L$), the quantity $\sqrt{2 m (E - V_0)}$
 is imaginary and leads to 
\begin{equation}
    \label{tempoclassicoquantico}
    \left< \mathbb{T} \right> (x) = i x m \frac{\int \mathrm{d} E \,\left[2 m (V_0 - E) \right]^{-1/2}  G(E,x)^2}{ \int \mathrm{d} E \, G(E,x) ^2}.
\end{equation}
We apply the previous expression to the limits $0$ and $E_{\text{\tiny max}}$ (satisfying the strong potential approximation), and make the substitution $u=\sqrt{2m(V_0-E)}$, $\Longrightarrow \mathrm{d}u = -m/\sqrt{2m(V_0-E)}\,\mathrm{d}E$. The numerator becomes
\begin{eqnarray}
i x m \int_0^{E_{\text{\tiny max}}} \mathrm{d} E \frac{G(E,x)^2}{p_E} &=& -ix\int_{p_0}^{p_E} \mathrm{d}u\, e^{\left(-\frac{2}{\hbar}ux\right)} \nonumber \\
  &=& \frac{i\hbar}{2}\left[ e^{-\frac{2}{\hbar}p_Ex}-e^{-\frac{2}{\hbar}p_0 x}\right], \nonumber\\
\end{eqnarray}
with $p_0 = \sqrt{2 m V_0}$ and $p_E = \sqrt{2 m (V_0 - E_{\text{\tiny max}})}$. The denominator follows from the same substitution, giving
\begin{eqnarray}
  \int_0^{E_{\text{\tiny max}}} \mathrm{d}E \, e^{- \frac{2}{\hbar} p_E x}  &=&  +\frac{ e^{- \frac{2}{\hbar}p_E x }\hbar \left( \hbar + 2 p_E x\right) }{4 m x^2} \nonumber\\
  && - \frac{e^{- \frac{2 p_0 x}{\hbar}} \hbar \left( \hbar + 2 p_0 x \right)}{4 m x^2}.
\end{eqnarray}
Together, both equations give us our main result:
%\begin{widetext} 
\begin{eqnarray}
  \label{timetravelan}
 T_{\text{\tiny STS}}(0 \to L) &=& \left< \mathbb{T} \right> (L) - \left< \mathbb{T} \right> (0)\nonumber\\
 & & \cr
 &=&  \frac{2imL^2\left[ 1 - \gamma  \right]}{\hbar \left[ 1 - \gamma \right] + 2L\left[p_E-p_0 \gamma \right]},
\end{eqnarray}
where we used $\lim_{x \to 0}  \left< \mathbb{T} \right>(x) = 0$, and $\gamma = e^{-\frac{2}{\hbar}\left(p_0-p_E\right)L }$.  In the classical limit $\hbar\to 0$ we obtain 
\begin{equation}
T^{\text{\tiny (class)}}_{\text{\tiny STS}}(0 \to L)  = \frac{imL}{p_E},
\label{Tclas}
\end{equation}
while for $E_{\text{\tiny max}}\to 0$ (no wave packet)
\begin{equation}
T^{(E_{\text{\tiny max}}=0)}_{\text{\tiny STS}}(0 \to L) =\frac{2imL^2}{\hbar+2Lp_0}.
\end{equation}
In both limits $E_{\text{\tiny max}}\to 0$ and $\hbar\to 0$, we have $T_{\text{\tiny STS}}(0 \to L) = imL/p_0$.

\begin{figure*}[t]
     \centering 
     \begin{subfigure}{0.33\textwidth}
         \centering
         \includegraphics[width=\textwidth]{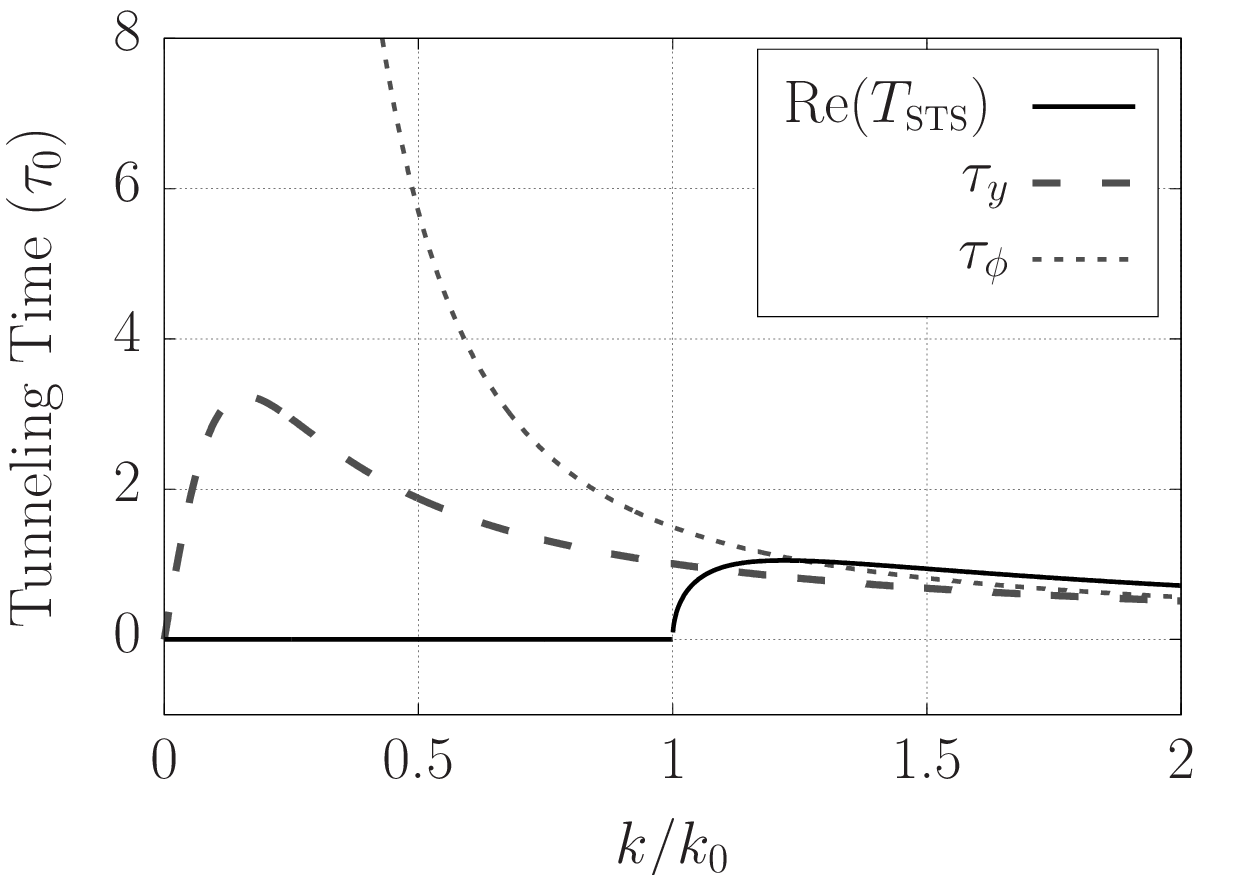}
         \vspace{-\baselineskip}
         \caption{$k_0 L = \pi/10$}
         \label{weakreal}
     \end{subfigure}%\hfill
     \begin{subfigure}{0.33\textwidth}
         \centering
         \includegraphics[width=\textwidth]{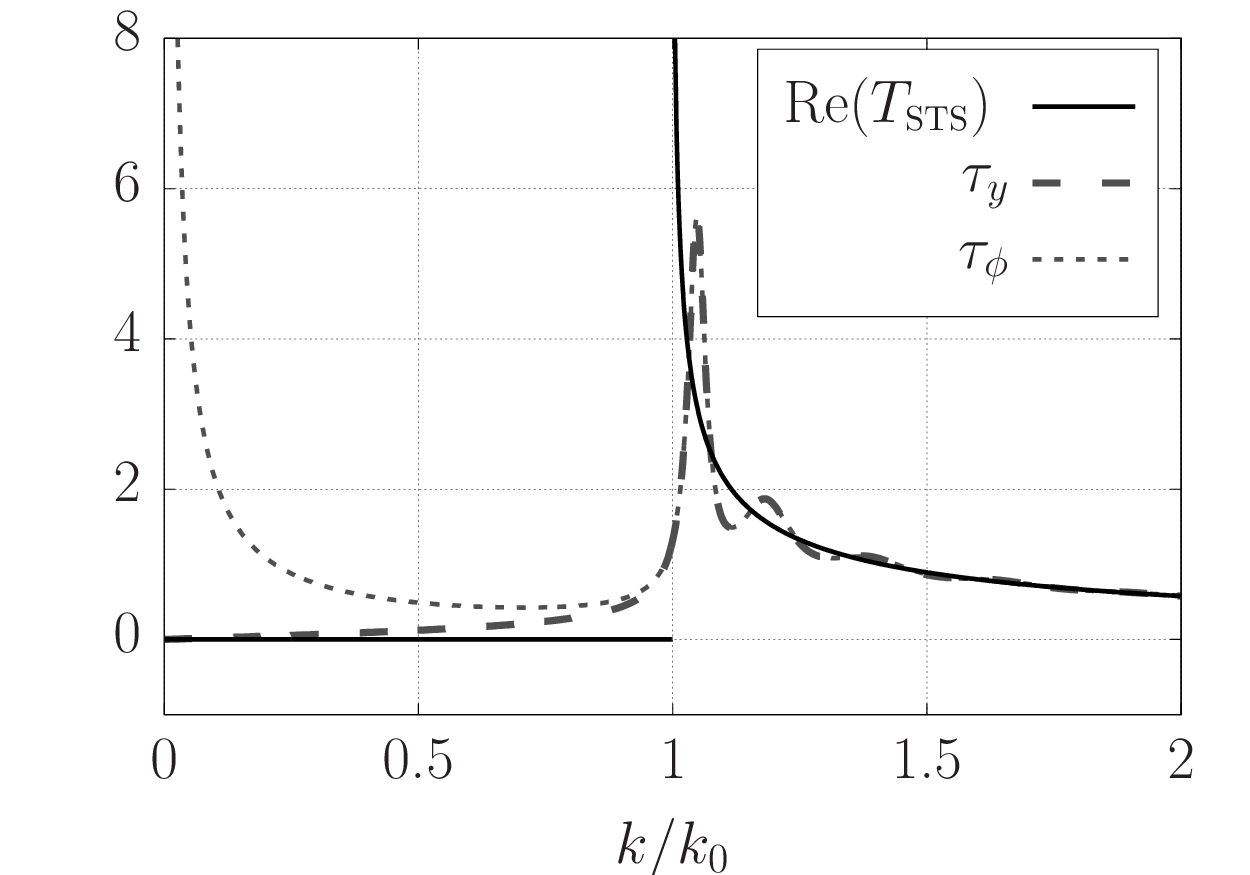}
         \vspace{-\baselineskip}
         \caption{$k_0 L = 3 \pi$}
         \label{medreal}
     \end{subfigure}
          \begin{subfigure}{0.33\textwidth}
         \centering
         \includegraphics[width=\textwidth]{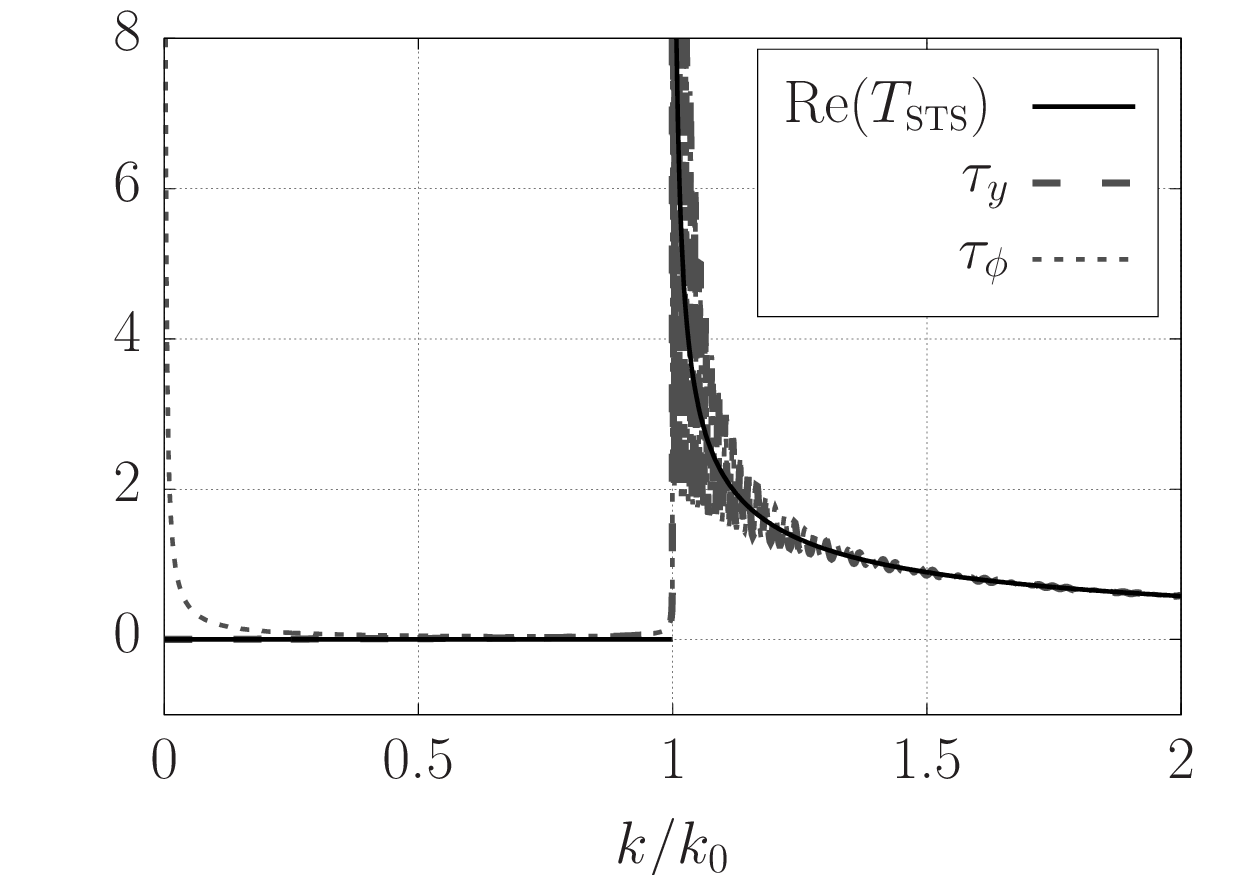}
         \vspace{-\baselineskip}
         \caption{$k_0 L = 30 \pi$}
         \label{strongreal}         
     \end{subfigure}
     \hfill
     \begin{subfigure}{0.33\textwidth}
                \centering
         \includegraphics[width=\textwidth]{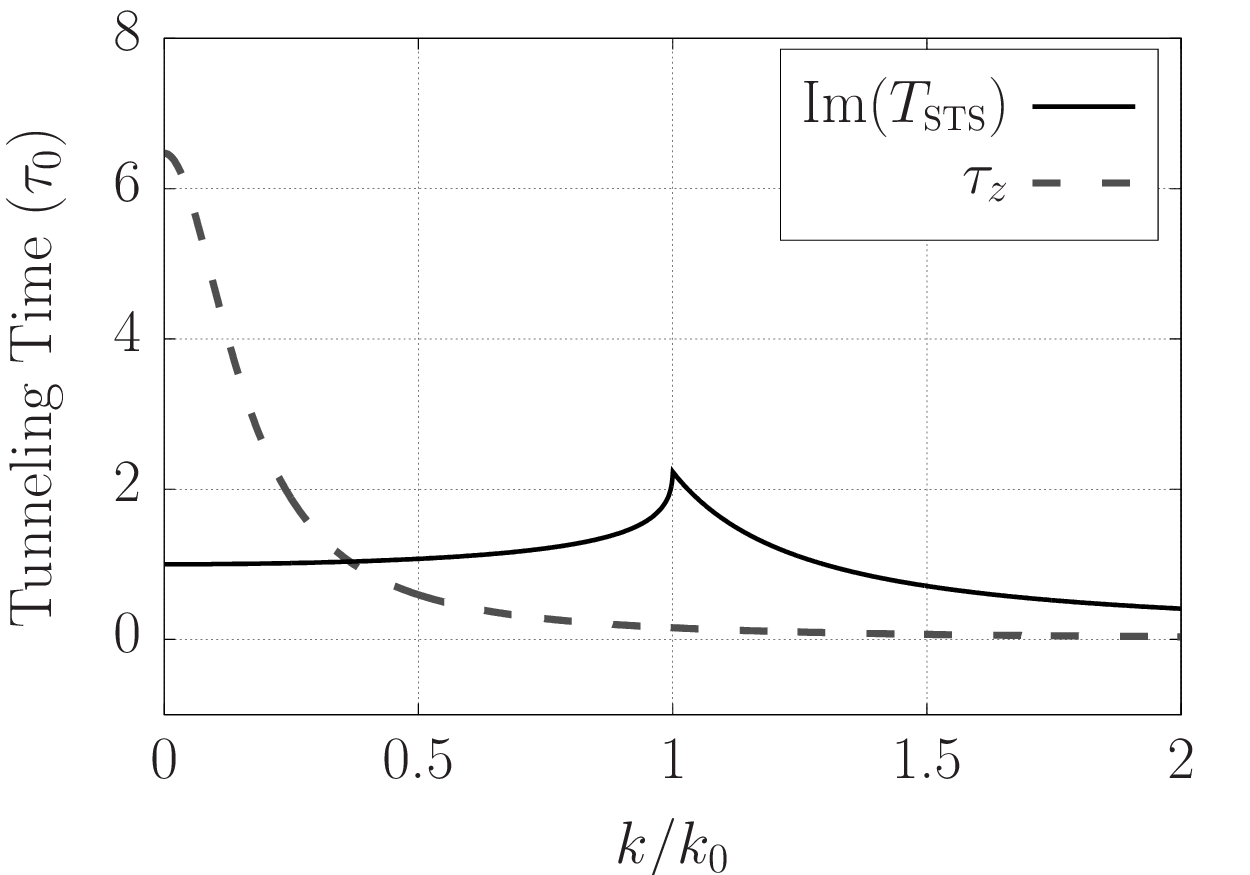}
         \vspace{-\baselineskip}
         \caption{$k_0 L = \pi/10$}
         \label{weakimag}
     \end{subfigure}%\hfill
     \begin{subfigure}{0.33\textwidth}
         \centering
         \includegraphics[width=\textwidth]{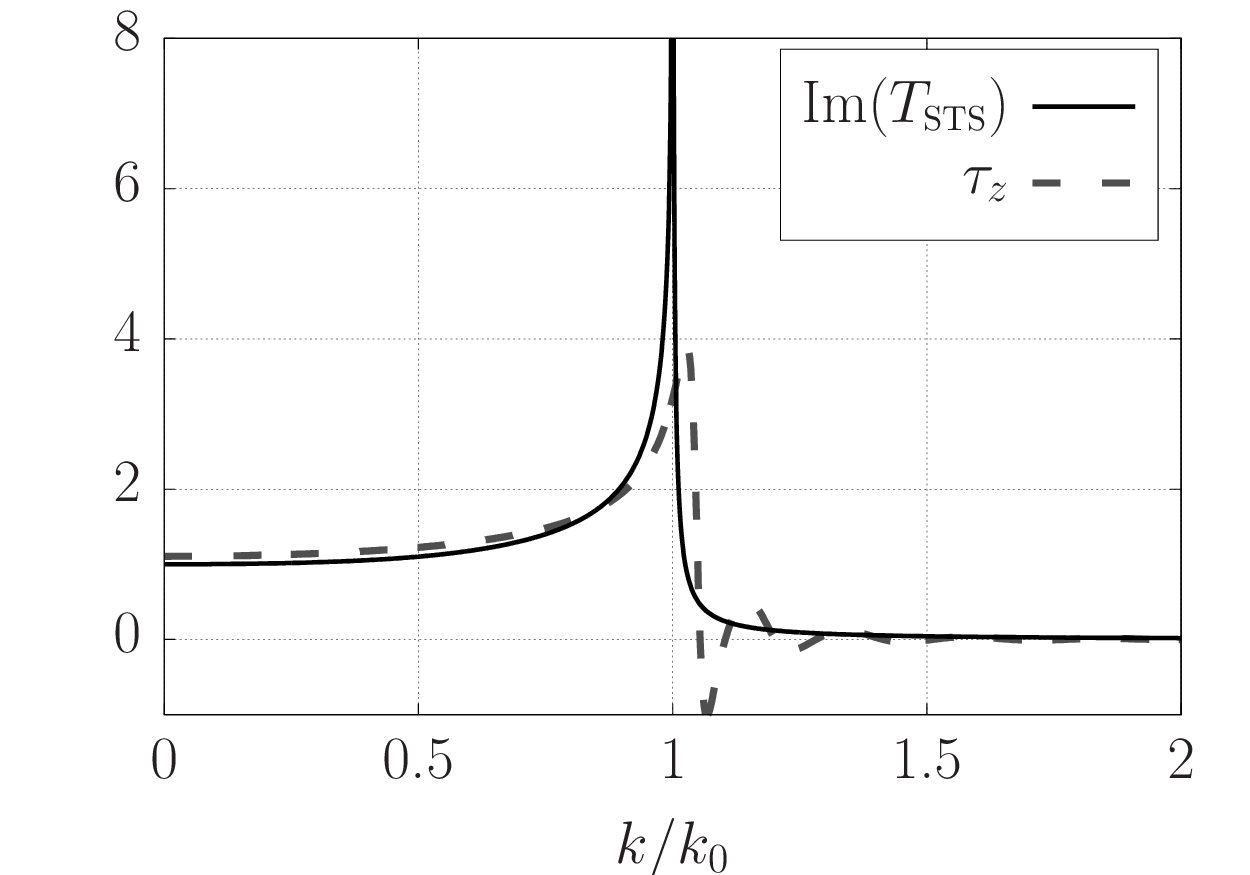}
         \vspace{-\baselineskip}
         \caption{$k_0 L = 3 \pi$}
         \label{medimag}
     \end{subfigure}
          \begin{subfigure}{0.33\textwidth}
         \centering
         \includegraphics[width=\textwidth]{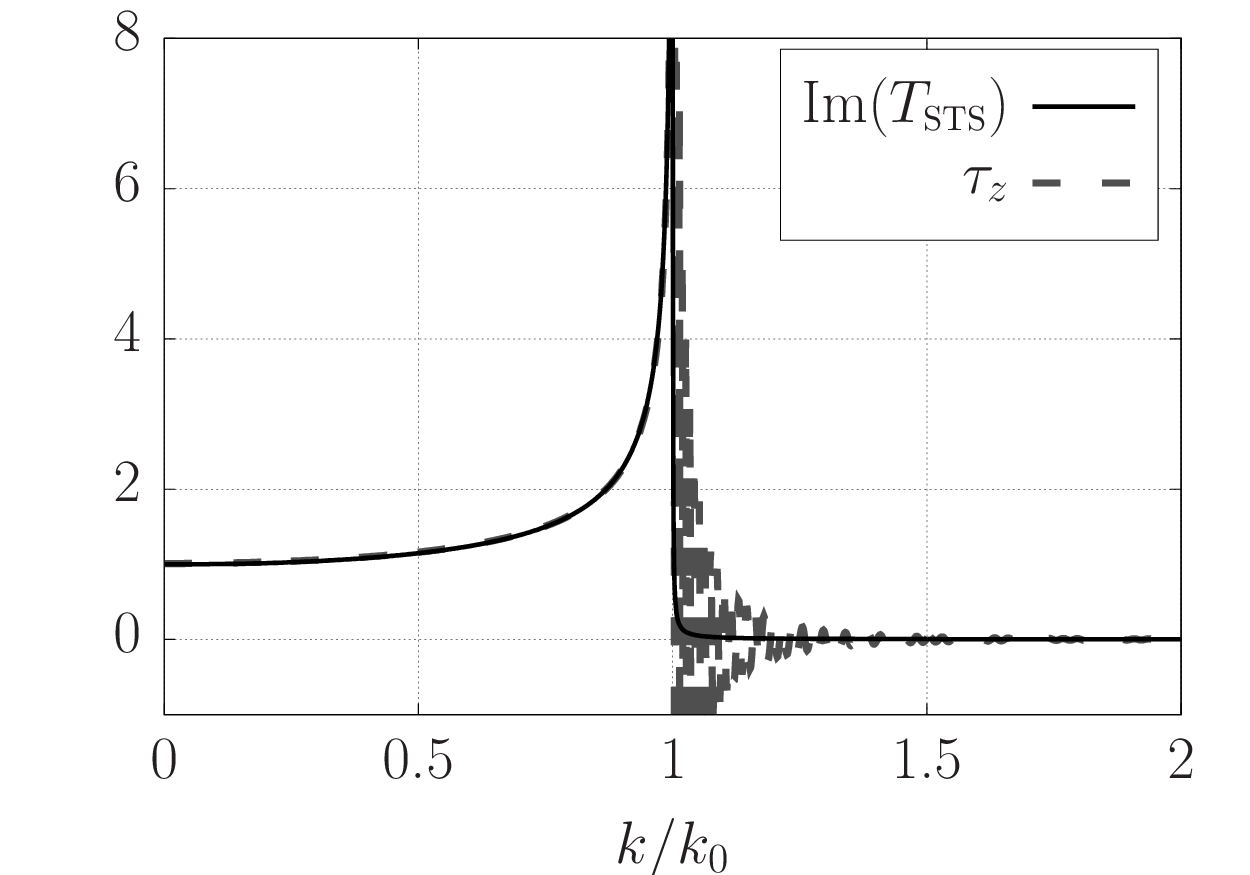}
         \vspace{-\baselineskip}
         \caption{$k_0 L = 30 \pi$}
         \label{strongimag}
     \end{subfigure}
     \caption{Comparison between Eq.~(\ref{timetravelan}) and the travel times $\tau_z$, $\tau_y = \tau_{\text{\tiny D}}$ and $\tau_{\phi}$, as obtained in Ref.~{\cite{BL3}}. Here, we have $k_0 = \sqrt{2m V_0}/\hbar$, $k = \sqrt{2 m E_{\text{\tiny max}}}/\hbar$ for $T_{\text{\tiny STS}}$ and $k = \sqrt{2 m E}/\hbar$ for the other times. $k_0 L$ gives us the strength of the barrier, while $\tau_0 = m L/\hbar k_0$ gives us a characteristic time for the barrier. We notice that, for a weak barrier (Figs.~\ref{weakreal} and~\ref{weakimag}), our result differs greatly. As we increase the barrier strength, $T_{\text{\tiny STS}}$ starts to agree more for large $k$ while acting as an average for $k > k_0$.}
        \label{comparison}
\end{figure*}

Equation (\ref{timetravelan}) can be compared to the characteristic times obtained from the precession of spin in an infinitesimal field in the $\hat{z}$ direction (for more details, check further Ref.~\cite{BL3}), also known as Larmor times $\tau_z$, $\tau_y$ (which coincides with the dwell time $\tau_{\text{\tiny D}}$) and the phase time $\tau_\phi$~\cite{hauge89}. The tunnelling time, in units of the characteristic barrier time $\tau_0=mL/\hbar k_0$,  is shown in Fig.~\ref{comparison} as a function of $k/k_0$.
For $k/k_0<1$, we have energies below the barrier, and for $k/k_0>1$, energies above the barrier. Distinct strengths of the barrier $k_0L\equiv p_0 L /\hbar$ are used. Top row of Fig.~\ref{comparison} compares the real part of $T_{\text{\tiny STS}}(0 \to L)$ with $\tau_y, \tau_{\phi}$, and bottom row,  the imaginary part of  $T_{\text{\tiny STS}}(0 \to L)$ with $\tau_z$. For small strengths, $k_0L=\pi/10$,   Figs.~\ref{weakreal} and~\ref{weakimag} show that, besides for some larger values of $k/k_0\gtrsim 1.5$, curves do not match at all. This is expected since Eq.~(\ref{timetravelan}) is obtained from a strong potential approximation. For stronger barriers, we observe that the imaginary part of Eq.~(\ref{timetravelan}) starts to have a good agreement with $\tau_z$ inside the barrier while behaving as some average of the oscillations outside, as shown in Figs.~\ref{medimag} and~\ref{strongimag}.  We also observe that $\mbox{Im} (T_{\text{\tiny STS}}(0 \to L))$ always begins at $\tau_0 = m L/\hbar k_0=m L/p_0$, which is the imaginary part of $T_{\text{\tiny STS}}(0 \to L)$ for $E_{\text{\tiny max}} \to 0$ and $\hbar \to 0$. For the real part, on the other hand, since Eq.~(\ref{timetravelan}) is imaginary for $k < k_0$, it always vanishes inside the barrier. Outside the barrier, as shown in Figs.~\ref{medreal} and~\ref{strongreal}, we recognize that the real part is again an average of the times $\tau_y$ and $\tau_\phi$, which coincide for $k > k_0$ and oscillates very rapidly for stronger barriers.
We also notice that, for increasing $k$, $\mbox{Im}(T_{\text{\tiny STS}}(0 \to L))$ decays very fast, goint to $0$ almost immediatly for stronger barriers, while $\mbox{Re}(T_{\text{\tiny STS}}(0 \to L))$ remains considerable. Remembering that our findings come from expectation values of the operator  $\mathbb{T}$, it may imply that the presence of the barrier makes $\mathbb{T}$ act like a Hermitian operator (with only real eigenvalues) for energies above the barrier or an anti-Hermitian operator (with only complex eigenvalues) for energies below the barrier. Worth noticing also that since Eq.~(\ref{integralnumeradorcomplexo}) has the real $\sqrt{2 m (E - V_0)}$ condition included, there is no problem in applying Eq.~(\ref{timetravelan}) for energies outside the barrier. 

Though Eq.~(\ref{timetravelan}) is already obtained from a Taylor expansion, we may use further approximations to understand the underlying physical properties of the tunnelling time: up to second order in $E_{\text{\tiny max}}/V_0$, we have
  \begin{eqnarray}
    \label{timetravelexpanded}
    T_{\text{\tiny STS}}(0 \to L) &\simeq& \frac{ i m L}{p_0} \left(1 + \frac{E_{\text{\tiny max}}}{4 V_0} \right) \nonumber\\
    && + \left( \frac{ i m L}{8 p_0} + \frac{ i m L^2}{24 \hbar} \right)\left(\frac{E_{\text{\tiny max}}}{V_0} \right)^2.
  \end{eqnarray}
The first two terms in this expansion, the first line of Eq.~(\ref{timetravelexpanded}), can be rewritten approximately as
\begin{equation}
  \label{tunnelingtime2ndorder}
 \frac{ i m L}{p_0} \left(1 + \frac{E_{\text{\tiny max}}}{4 V_0} \right) \simeq \frac{i m L}{\sqrt{2 m \left(V_0 - E_{\text{\tiny max}}/2 \right)}}
\end{equation}
One can understand this result in two ways: firstly, this is the first-order expansion of the time of travel $m L/p$ of a classical particle with energy $E_{\text{\tiny max}}/2$ and momentum $p = \sqrt{2m (V_0 - E_{\text{\tiny max}}/2)}$ inside the barrier (except for a multiplicative $i$).  Note that $E_{\text{\tiny max}}/2$ is the mean energy between $0$ and $E_{\text{\tiny max}}$.

\begin{figure}[b]
    \centering
         \includegraphics[width=0.4\textwidth]{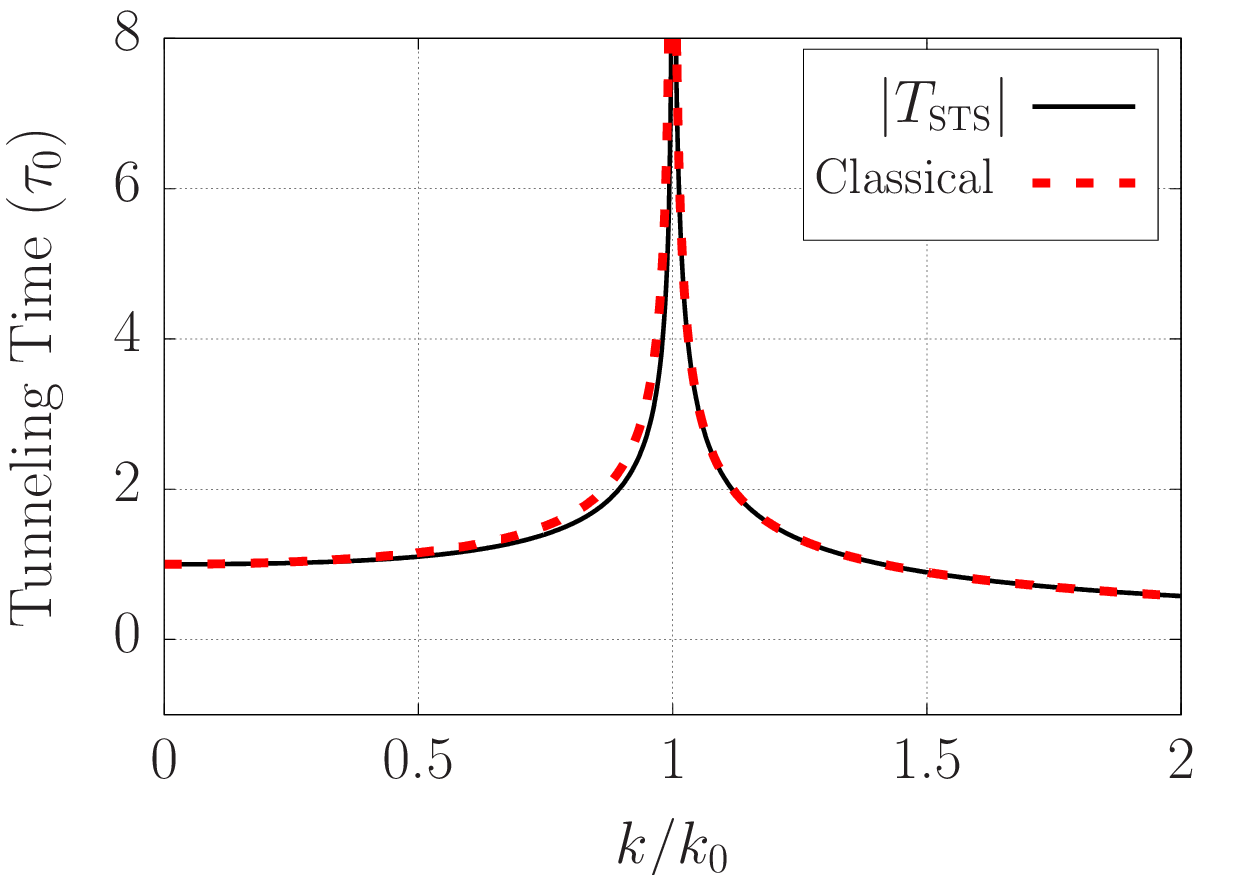}
%         \vspace{0.01\baselineskip}
         \caption{Comparisons between Eq.~(\ref{timetravelan}) in absolute values and the ``classical'' time $\left|m L/\sqrt{2 m (E_{\text{\tiny max}} - V_0)}\right|$ for $k_0 L = 3 \pi$, in units of $\tau_0 = m L/\hbar k_0$.}
         \label{classvsours}
  \end{figure}

Secondly, when we expand the momentum $\sqrt{2 m (V_0 - E)}$, we obtain the same result through the energy average of time-of-arrival of classical particles, except for a multiplicative $i$:
  \begin{eqnarray}
    \left<t \right>_{\text{\tiny class}} &\equiv& \frac{1}{E_{\text{\tiny max}}} \int_0^{E_{\text{\tiny max}}} \mathrm{d} E\, \frac{m L}{\sqrt{2 m (V_0 - E)}} \nonumber\\
    &\simeq& \frac{m L}{E_{\text{\tiny max}} p_0} \int_0^{E_{\text{\tiny max}}} \mathrm{d} E \, \left(1 + \frac{E}{2 V_0}\right) \nonumber\\
    &=& \frac{m L}{p_0} \left(1 + \frac{ E_{\text{\tiny max}}}{4 V_0} \right)\nonumber \\
    &\simeq& -i T_{\text{\tiny STS}}\left(0 \to L \right).
  \end{eqnarray}
We see that the time the particle takes to travel through the barrier is a classical-like contribution plus a quantum correction. We compare Eq.~(\ref{timetravelan}) with a classical particle with energy $E_{\text{\tiny max}}$ and  time of travel $ m L/\sqrt{2 m (E_{\text{\tiny max}} - V_0)}$ in Fig.~\ref{classvsours}. We see that outside the barrier, our result agrees very well with the classical time and is slightly different near $k \equiv \sqrt{2 m E_{\text{\tiny max}}} = k_0 \equiv \sqrt{2 m V_0}$, as expected since Eq.~(\ref{timetravelan}) is obtained through a strong potential approximation. Thus, our results signal that classical time is the most probable time.

  The travelling time from Eqs.~(\ref{travelLto0}), (\ref{timetravelan}) and~(\ref{timetravelexpanded}) can also be compared to other approximated tunnelling times expressions, which are summarized in Table I. Agreements are found in the limit $V_0\gg E$ with the Larmor time (except for the factor $i$) and the complex time (except for the signal). The agreement is also obtained compared to the imaginary part of the tunnelling time $\tau_{\mbox{\tiny S}}$ obtained in the telegrapher's equation. In this case, an additional term proportional to $(mL/p)^2$ is observed, where $a$ is the friction coefficient due to dissipation, that may be related to the second line in Eq.~(\ref{timetravelexpanded}): one writes
      \begin{equation}
        \frac{i m L^2}{24 \hbar} \left(\frac{E_{\text{\tiny max}}}{V_0} \right)^2 = \frac{i m^2 E_{\text{\tiny max}}^2}{ 12 \hbar V_0} \left( \frac{L}{p_0} \right)^2,
      \end{equation}
in the strong potential limit (we discarded the term proportional to $E_{\text{\tiny max}}^2/V_0^3$), and identity $a = i m^2 E_{\text{\tiny max}}^2/12 \hbar V_0$.

 Not shown in Fig.~\ref{comparison} is the Büttiker-Landauer time~\cite{BL3, hauge89} $\tau_{\text{\tiny B-L}} \equiv \tau_x = \sqrt{\tau_z^2 + \tau_y^2}$, but we can see that, for large $k_0 L$, only $\tau_z$ is important for $k<k_0$, while only $\tau_y$ contributes for $k>k_0$, meaning that our results agree in such limits.
 \begin{table}
\caption{Specific tunneling times expressions through a rectangular box with width $L$,  $\kappa^2=k_0^2-k^2$,  with $k$ and $k_0$ as given in Fig.~\ref{comparison}.  The parameter $a$ is related to the friction coefficient that enters the telegrapher's equation, and $v = p/m$ is the particle's velocity through the barrier. Extracted from Refs.~\cite{hauge89,secao122,diasparisio}. } 
\begin{tabular}{|c|c|c|}
\hline
Quantity        &      Expression          & Limit $V_0\gg E$ \\ \hline
Phase time    &    $\tau_{\mbox{\tiny P}} = \tau_{\phi}\simeq\frac{2m}{\hbar k\kappa}$ & $\tau_{\mbox{\tiny P}}\to\frac{2m}{\hbar k k_0}$\\ \hline
Dwell time    &  $\tau_{\mbox{\tiny D}} = \tau_y\simeq\frac{2mk}{\hbar\kappa k_0^2} $ & $\tau_{\mbox{\tiny D}}\to 0 $ \\ \hline
Larmor time & $\tau_{\mbox{\tiny L}} = \tau_{z}\simeq \frac{mL}{\hbar\kappa}$  & $\tau_{\mbox{\tiny L}}\to \frac{mL}{\hbar k_0}$\\  \hline
Complex time & $\mbox{Im}[\tau_{\mbox{\tiny C}}]=-\tau_{\mbox{\tiny L}}$  & $\mbox{Im}[\tau_{\mbox{\tiny C}}]\to -\frac{mL}{\hbar k_0}$\\  \hline
Stochastic model &  $\tau_{\mbox{\tiny S}} \simeq a \left(\frac{mL}{\hbar \kappa}  \right)^2 + i \frac{mL}{\hbar \kappa}$ &  $\frac{imL}{\hbar \kappa}$ \\ \hline
STS model  &  Eq.~(\ref{timetravelan}) &  $i mL / \hbar k_0$ \\ \hline
\end{tabular}
\end{table}

%==================================================================================================================

\subsection{Application of travelling time: a constant distribution for a wave packet moving to the right in a weak potential}
In the case of a weak potential, we apply Eqs.~(\ref{ansatzpotencialfraco}) and~(\ref{energiapotencialfraco}) to Eq.~(\ref{valoresperadofinal}) to obtain the travelling time for this situation. Similarly to Sec.~\ref{applicationstrong}, we consider a wave packet with $C_E^- = 0$ and $C_E^+ = C = \text{const}$. But, contrary to Eq.~(\ref{tempoclassicoquantico}), $G(E,x)$ is a complex exponential $\left|G(E,x) \right|^2 = 1$, leading us to
\begin{eqnarray}
  \left< \mathbb{T} \right> (x) &=&   \frac{\int_{E_i}^{E_f} \mathrm{d} E \, x m \left[2 m (V_0 - E) \right]^{-1/2}}{\int_{E_i}^{E_f} \mathrm{d} E } \nonumber\\
  &=& \frac{\int_{E_i}^{E_f} \mathrm{d} E \, x m p_E^{-1}}{\int_{E_i}^{E_f} \mathrm{d} E },
  \end{eqnarray}
$E_f$ and $E_i$ are such as to allow us to use the weak potential approximation of Sec.~\ref{weakpotential}, and $p_E = \sqrt{2m (E - V_0)}$ Then, the time of travelling is written as 
\begin{eqnarray}
  \label{tempoweakpotential}
  T_{\text{\tiny travel}}(0 \to L) &=& \left< \mathbb{T} \right>(L) - \left< \mathbb{T} \right> (0), \nonumber\\
%  &=& \frac{1}{\int_{E_i}^{E_f} \mathrm{d} E } \int_{E_i}^{E_f} \mathrm{d} E \, \frac{mL }{ p_E} \nonumber\\
  &=&  \frac{1}{E_f - E_i} \int_{E_i}^{E_f} \mathrm{d} E \, \frac{mL }{ p_E}, \nonumber\\
  &=& \frac{L}{\Delta E} \left[\sqrt{2 m (E_f - V_0)} - \sqrt{2 m (E_i - V_0)} \right],\nonumber\\
\end{eqnarray}
with $\Delta E = E_f - E_i$ and $\left< \mathbb{T} \right> (0) = 0$. Equation (\ref{tempoweakpotential}) says that if a particle has an energy greater than the barrier, the expected time of arrival is an average (in the energies) of the \emph{classical} time-of-arrival $\tau_{\text{\tiny C}} = L m/p_E$.

For the free particle, $V_0 = 0$, we then have 
  \begin{equation}
    \label{tempoweakpotentialfree}
    T_{\text{\tiny travel}}(0 \to L) = \frac{L}{\Delta E} \left[\sqrt{2 m E_f} - \sqrt{2 m E_i} \right].
    \end{equation}
  Consider the case in which $E_i = 0$ and $E_f = E_{\text{\tiny max}}$, the same energies as the tunnelling case in Sec.~\ref{applicationstrong}. We obtain
    \begin{equation}
      \label{classicalweak0emax}
      T^{\text{\tiny free}}_{\text{\tiny travel}}(0 \to L) = 2\frac{m L}{\sqrt{2m E_{\text{\tiny max}}}}.
  \end{equation}
  We then compare Eq.~(\ref{classicalweak0emax}) to the absolute value of the approximated tunnelling time, given by Eq.~(\ref{tunnelingtime2ndorder}):
  \begin{eqnarray}
    \frac{T_{\text{\tiny travel}}^{\text{\tiny free}}}{T_{\text{\tiny travel}}}&=&  2 \frac{ m L}{\sqrt{2m E_{\text{\tiny max}}}}  \left( \frac{ m L}{\sqrt{2m (V_0 - E_{\text{\tiny max}})}} \right)^{-1} \nonumber\\
    &=&  2 \sqrt{\frac{V_0 - E_{\text{\tiny max}}}{E_{\text{\tiny max}}}} \nonumber\\
    &\simeq& 2 \sqrt{V_0 / E_{\text{\tiny max}}} \gg 1,
  \end{eqnarray}
  where we used the fact that $V_0 \gg E_{\text{\tiny max}}$. This is compatible with known results from~\cite{winful2006} and references therein, where the tunnelling time is shorter than the time a free particle would take to cross the same region.

%==================================================================================================================
\section{\label{conclusion}Final remarks}
This work summarises the main ideas of a recently proposed formalism that tries to include and understand a time operator in QM. The formalism is spacetime-symmetric (STS) and allows for predicting times-of-flight and tunnelling times. We apply the formalism for a particle with energy $E$ under weak and strong constant potentials, namely a rectangular barrier with length $L$ and intensity $V_0$. Connections formulas between distinct regions of motion are provided to obtain an explicit expression for the tunnelling time through a barrier. Using a wave packet with a constant distribution of energies, we show that the tunnelling time in the STS formalism is in agreement with previous times, such as $\tau_z$, $\tau_y = \tau_{\text{\tiny D}}$, $\tau_{\phi}$ from Ref.\cite{BL3}. Furthermore, we provide, in first order, the average of classical times-of-flight for an ensemble of particles with momenta $\sqrt{2m(V_0 - E)}$, except for a complex multiplicative unit. The appearance of an imaginary time is consistent with a similar mechanism to the one that appears in Büttiker's model using a Larmor clock~\cite{BL3, secao122}, which has a real part $\tau_{\phi}$ and an imaginary part $\tau_z$, possibly bringing to light the discussion of (anti-)hermiticity of a time operator for the tunnelling case. Our procedure differs from those obtained in Ref.~\cite{diasparisio} when considering the connection formulas between allowed and prohibited regions, allowing us to furnish analytical expressions for the tunnelling times.

The STS formalism is promising. For example, apart from helping our general understanding of the time in QM, it could assist in using fractional derivatives and integrals in physics and their interpretations~\cite{laskin, naber, stanislavsky, laskin1, laskin2, laskin3}. Especially we can see, when comparing the dynamical equations for the weak \textit{versus} strong potentials, that the order of the time derivative varies from $1/2$ to $1$, respectively.  Besides, it could motivate further studies giving more insights into the symmetries between spacetime and energy-momentum.

Generally speaking, solving  Eq.~(\ref{eqprojetada}) is the main challenge. One possible way to do it is using the Fourier transform of the square-root operator. In Ref.~\cite{samkokilbas}, Sec. 28.2 gives us the treatment for powers of the operator $- \Delta x + \partial_t$, but for different integrodifferential operators. In principle, this could be expanded to the Momentum operator in the STS formalism and give us solutions beyond the scope of constant potentials. We could then compare predictions with, for instance, the toy model for the Stark problem of Ref.~\cite{yusofsani}. Possible problems of the inverse Fourier transform convergence could be avoided by limiting the integration frequencies, the barrier acting as a filter, as justified by Eq.~(12.5) of Ref~\cite{secao122}.

%==================================================================================================================
\begin{acknowledgments}
  This study was financed in part by the Coordenação de Aperfeiçoamento de Pessoal de Nível Superior - Brasil (CAPES) – Finance Code  001. MWB thanks CNPq (Brazilian agency) for financial support (Grant Numbers.~ 310792/2018-5).
\end{acknowledgments}

\providecommand{\noopsort}[1]{}\providecommand{\singleletter}[1]{#1}%
%

%==================================================================================================================

%\bibliography{larabeims}% Produces the bibliography via BibTeX.

\end{document}